# ELECTROMAGNETIC THEORY OF SOUND AND PHONONS IN LIQUID

A.A. Stupka

Oles Honchar Dnipropetrovsk National University, Department of Quantum Macrophysics,
Gagarin Ave. 72, 49010, Dnipropetrovsk, Ukraine
e-mail: antonstupka@mail.ru

*Abstract* - **Sound waves in and dielectric liquid, that consists of subsystems of valency electrons with effective mass and ions which interact through the long-wave potential electric field are considered, on analogies with and metallic liquid, but with the requirement, that electric current is absent. It is shown that usual description of sound wave in an ideal liquid as wave of mass density and mass velocity follows from the offered consideration. It is also shown that sound waves with other side can be considered as waves of potential electric field in an environment. Introduction of phonons in both cases is considered.**

I. INTRODUCTION

Sound or acoustic waves are the waves, existence of which is conditioned by resilient forces which arise up during deformation of environment ([1] p. 34). Phenomenologic hydrodynamic consideration for long waves in a liquid environment is well known [2]: there is oscillation of mass density $\rho$ and mass velocity $\mathbf{v}$. But nature of the mentioned resilient forces obviously not only quantum but also electromagnetic and we make attempt show the role of the long-wave electric field in a sound wave.

We consider distribution of sound wave in a liquid, as continuous environment. For simplicity we ignore all dissipation processes. It is well known [3] for liquid metals, that electronic and ionic subsystems demonstrate a different behavior. Velocity of sound $u_s$ is considerably bigger, than thermal rate of movement of subsystem of ions and we can ignore such motion. But for an electronic subsystem sound is a slow process, because Fermi-velocity considerably bigger from $u_s$. Sound waves in metals can be considered as ion-sound waves in plasma [3,4]. For dielectric liquids correlations between characteristic velocities are the same. Characteristic velocity of ions can be appraised as $u_i \sim \sqrt{T/m_i}$, where $T$ is the temperature of liquid, $m_i$ is mass of ion (identical ions). Electrons in general case form Fermi-liquid with some function of pressure, however for quality estimation after the order of value it is possible to use approximation of degenerated Fermi-gas with a zero temperature and pressure [5] $P_e = (3\pi^2)^{2/3}\hbar^2 n_e^{5/3}/5m_e$, that gives $u_e^2 = (3\pi^2)^{2/3}\hbar^2 n_e^{2/3}/3m_e^2 = v_{Fe}^2/3$, where $v_{Fe}^2$ is the square of electronic Fermi-velocity, $n_e$ is density of valency electrons. So, in the system there are three characteristic velocities $u_e \gg u_s \gg u_i$. This fact suggests an idea to consider valency electrons and ions of dielectric liquid in the process of distribution of sound wave as two subsystems which interacts. But unlike the case of metal, density of electric current is to be zero

$$\mathbf{j}_i + \mathbf{j}_e = 0, \quad (1)$$

where $\mathbf{j}_a$ is density of electric current of components $a = i, e$. So, in our macroscopic consideration a condition (1) means that environment can be a dielectric. In the hydrodynamic approximation current components is [6] $\mathbf{j}_a = e_a n_a \mathbf{v}_a$, where $e_a$ is the proper charge. For each components it is possible to write down continuity equation

$$\partial n_i/\partial t + div n_i \mathbf{v}_i = 0, \quad \partial n_e/\partial t + div n_e \mathbf{v}_e = 0 \quad (2)$$

and Euler equation: for ionic components it is possible to ignore pressure because we consider a high-frequency process in relation to this component

$$\rho_i \partial v_{i\alpha}/\partial t = -\rho_i(v_{i\beta}\nabla_\beta)v_{i\alpha} + Zen_i\left(E_\alpha + [\mathbf{v}_i, \mathbf{B}]_\alpha/c\right), \quad (3)$$

for electronic component pressure $P_e$ is determined by concrete properties of energy of interaction of electron with the ion, that means by the matter. At our consideration $P_e$ we consider the known function

$$\rho_e \partial v_{e\alpha}/\partial t = -\rho_e(v_{e\beta}\nabla_\beta)v_{e\alpha} - \partial P_e/\partial x_\alpha - en_e\left(E_\alpha + [\mathbf{v}_e,\mathbf{B}]_\alpha/c\right), \tag{4}$$

where $\rho_a = m_a n_a$ is density of mass. For quality consideration there is possible the use of expression for pressure of degenerated ideal Fermi-gas for electronic components. Usually, we must enter effective mass of electronic component in approximation of the strongly coupled electrons ([7] c. 136), because for dielectrics in the first approximation it is possible to neglect covering of wave functions of electrons which behave to different ions. Like enters effective mass at consideration of not alkaline metals for transition to approximation of free quasielectrons and subsequent study of processes of transfer of current electronic component. For a dielectric such processes do not exist, that is taken into account by the condition (1), but the subsystems of electrons and ions take part in collective motions with certain frequencies, in particular in sound oscillations, as separate subsystems. $E_\alpha$ selfconsistent long-wave electric field (short-wave part is included in effective mass), that describes interaction of the charged subsystems of environment at collective motions. Consequently, we study a nondissipative liquid near an equilibrium and potential motions on the basis of equations (1)-(4).

## II. Sound Waves

We consider the adiabatic sound waves of small amplitude in the system. For this purpose we perform linearization of equations (1)-(4). From dielectric condition (1) and equations (2) - (4) we have

$$\mathbf{v}_e - \mathbf{v}_i = 0,\ \partial n_i/\partial t + n_{i0}div\mathbf{v}_i = 0,\ \partial n_e/\partial t + n_{e0}div\mathbf{v}_e = 0, \tag{5}$$

$$\partial \mathbf{v}_i/\partial t = Ze\mathbf{E}/m_i,\ \partial \mathbf{v}_e/\partial t = -u_e^2 \nabla n_e/n_{e0} - e\mathbf{E}/m_e, \tag{6}$$

where $u_e^2 = (\partial P_e/\partial n_e)_S/m_e$ is characteristic velocity. Here comfortably to pass to Fourier-components by rule $A(x,t) = \int d^3k d\omega A(k,\omega)e^{ikx-i\omega t}/(2\pi)^4$. Systems of equations for longitudinal and transversal in relation to a wave vector oscillations are divided. For transversal (vortical) parts of vectors we have

$$\mathbf{v}_e^t - \mathbf{v}_i^t = 0,\ -i\omega\mathbf{v}_i^t = Ze\mathbf{E}^t/m_i,\ -i\omega\mathbf{v}_e^t = -e\mathbf{E}^t/m_e. \tag{7}$$

The system (7) has the banal decision only, that transversal long-wave oscillations of the mentioned values do not exist in a dielectric liquid. Then, we consider longitudinal (potential) oscillations. Equations (5)-(6) give

$$v_e^l - v_i^l = 0. \tag{8}$$

$$-i\omega n_i + in_{i0}kv_i^l = 0,\ -i\omega n_e + in_{e0}kv_e^l = 0 \tag{9}$$

$$-i\omega v_i^l = ZeE^l/m_i,\ -i\omega v_e^l = -u_e^2 ikn_e/n_{e0} - eE^l/m_e \tag{10}$$

We have the system from five linear homogeneous equations, which has the nontrivial solution, if determinant of the system (8)-(10) equals to zero:

$$-ei^3 k^2 u_e^2 Z\omega/m_i + ei^3\omega^3/m_e + eZi^3\omega^3/m_i = 0. \tag{11}$$

This equation of the third degree has three roots: $\omega = 0$ does not interest us, and other two roots

$$\omega = \pm k u_e \sqrt{Zm_e/(m_i + Zm_e)} \tag{12}$$

corresponds to the coupled oscillations of all five variables. Thus, how simply to see from a condition (8), there are longitudinal oscillations of mass velocity and mass density, because electrons and ions move together. As dependence (12) of frequency on a wave vector is linear, it is possible to assert that the found oscillations are a sound. And it is shown by us, that there is oscillation of the longitudinal electric field in such wave $E^l$. Velocity of sound from (12) is

$$u_s = u_e \sqrt{Zm_e/(m_i + Zm_e)}. \tag{13}$$

The resulted consideration of sound oscillations allows simplification in two ways: transition to mechanistic description through united density of mass and mass velocity [2] and description as oscillations of the longitudinal electric field with introduction of vector potential as conjugating coordinate (see, e.g., [8]).

We obtain at first usual hydrodynamic equations from (8)-(10). We enter united hydrodynamic velocity $\mathbf{v} = \mathbf{v}_e = \mathbf{v}_i$. General mass density of liquid is $\rho = nm = \rho_i + \rho_e = n_i m_i + n_e m_e$, where $m$ is mass of molecule of liquid, therefore continuity equation is sum of equations (9)

$$-i\omega\rho + i\rho_0 k v^l = 0. \tag{14}$$

By adding of equations (10) we get rid of the field $E^l$:

$$-i\omega\rho_0 v^l = -u_e^2 i k n_e m_e. \tag{15}$$

Through electro-neutrality of dielectric we have a condition $Zn_i - n_e = 0$, that allows to express by determination of general density of mass

$$n_e = nm/(m_i/Z + m_e). \tag{16}$$

We use equation (16), to get rid of electron density at (15):

$$-i\omega\rho_0 v^l = -\left(\partial P_e(n)/\partial n\right)_S ikn. \tag{17}$$

It is well known [2], that system of equations (14) and (17) has the solution as sound wave with velocity $u = \sqrt{\left(\partial P/\partial n\right)_S/m}$. With the use of relation (16) it is easy to see that it corresponds our result (13), that $u = u_s$. Energy in the sound wave of small amplitude is given by an integral [9]

$$E = \int\left(\rho_0 v^2/2 + u^2\rho^2/2\rho_0\right)d^3x. \tag{18}$$

Standard procedure of quantization of normal coordinates in Shrodinger presentation allows to express operators of hydrodynamic velocity potential $\nabla\varphi = \mathbf{v}^l$ and density through operators of elimination and birth of phonons $c_k$ and $c_k^+$

$$\varphi(x) = \Sigma_k \sqrt{\hbar u}\left(c_k + c_{-k}^+\right)e^{i\mathbf{kx}}/\sqrt{2V\rho_0 k}, \quad \rho(x) = \Sigma_k i\sqrt{\hbar\rho_0 k}\left(c_k - c_{-k}^+\right)e^{i\mathbf{kx}}/\sqrt{2Vu}. \tag{19}$$

Putting expressions (19) in a quantum case of (18) we obtain the standard form of Hamiltonian for sound oscillations

$$\hat{H} = \Sigma_k uk\hbar\left(c_k^+ c_k + 1/2\right) \tag{20}$$

From other side from equations (8)-(10) it is possible to obtain equation for the electric field. To that end we express from (10) ionic velocity through strength $i\omega Z e E^l/m_i\omega^2 = v_i^l$, and with the use of continuity equation for electrons from equation (10) $-i\omega e E^l/m_e\left(\omega^2 - u_e^2 k^2\right) = v_e^l$. We put this velocity expressions in equation (8)

$$-i\omega E^l/m_e\left(\omega^2 - u_e^2 k^2\right) = i\omega Z E^l/m_i\omega^2. \tag{21}$$

Got homogeneous equation for $E^l$ has the nonzero solution at satisfaction of dispersion equation, that equivalent to (11). As known, vector potential of the electromagnetic field in the Hamilton gage is related to strength [8]: $\partial\mathbf{A}/\partial t = -c\mathbf{E}$, where $c$ is velocity of light. And operator of energy in a sound wave in place of (18) in new variables it is possible to write down as

$$\hat{H}_f = \int d^3x \hat{E}(x)^{l2}/8\pi + \int d^3x \int d^3x' \omega^2(x-x')\hat{A}(x)^l \hat{A}(x')^l/8\pi c^2. \tag{22}$$

The general case of energy of electromagnetic wave of arbitrary frequency in an environment is considered in [10]. Counter-term to the second term in (22) has renormalized interaction of the field with an environment. In presentation of the second quantization we have expressions

$$A_n^l(x) = c\Sigma_k\left(c_k + c_{-k}^+\right)e^{i\mathbf{kx}}k_n\sqrt{2\pi\hbar/k^3 uV}, \quad E_n^l(x) = \Sigma_k i\left(c_k - c_{-k}^+\right)e^{i\mathbf{kx}}k_n\sqrt{2\pi\hbar u/kV}. \tag{23}$$

which after substitution in hamiltonian (22) give the expected expression (20) through operators of birth and elimination. And obviously, that the same process of sound oscillations has the same set of eigen coordinates, and consequently, operators of elimination and creation $c_k$ and $c_k^+$ in expressions (19) and in expressions (23) corresponds the same quasi-particles - phonons.

REFERENCES

[1] V.A. Krasilnikov, V.V. Krylov, *Introduction into physical acoustics* (Nauka, Moscow, 1984, in Russian).
[2] L.D. Landau, E.M. Lifshitz, *Hydrodynamics* (Nauka, Moscow, 1986, in Russian).
[3] D. Bohm, T. Staver, Phys.Rev., **84**, 836 (1951).
[4] D. Pines, *Elementary exitations in solids* (Mir, Moscow, 1965, in Russian).
[5] L.D. Landau, E.M. Lifshitz, *Statistical physics*, P. 1 (Nauka, Moscow, 1976, in Russian).
[6] *Electrodynamics of plasma*, ed. by A.I. Akhiezer (Nauka, Moscow, 1974, in Russian).
[7] A.S. Davydov, *Theory of solid* (Nauka, Moscow, 1976).

[8] A.I. Sokolovsky, A.A. Stupka, Z.Yu. Chelbaevsky, Ukrainian Journal of Physics. **55**, 20 (2010).
[9] E.M. Lifshitz, L.P. Pitaevsky, *Statistical physics*, P. 2 (Nauka, Moscow, 1978).
[10] A.I. Sokolovsky, A.A. Stupka, Proc. of 12-th International Conference MMET. Odessa. 262 (2008).